\documentclass[a4paper,twoside,10pt]{letter}
\usepackage{graphicx,saj,multicol,subeqnarray}

\def\papertype{Original Scientific Paper}

\begin{document}
\baselineskip=3.1truemm
\columnsep=.5truecm
\newenvironment{lefteqnarray}{\arraycolsep=0pt\begin{eqnarray}}
{\end{eqnarray}\protect\aftergroup\ignorespaces}
\newenvironment{lefteqnarray*}{\arraycolsep=0pt\begin{eqnarray*}}
{\end{eqnarray*}\protect\aftergroup\ignorespaces}
\newenvironment{leftsubeqnarray}{\arraycolsep=0pt\begin{subeqnarray}}
{\end{subeqnarray}\protect\aftergroup\ignorespaces}
%

% Running titles

\markboth{\eightrm Long-term optical photometric monitoring of the FUor star V900 Mon} {\eightrm E. SEMKOV, S. PENEVA AND S. IBRYAMOV}

{\ }

\publ

\type

{\ }

% Title

\title{Long-term optical photometric monitoring of the FUor star V900 Mon}

% Authors

\authors{E. H. Semkov$^{1}$, S. P. Peneva$^{1}$ and S. I. Ibryamov$^{2}$}

\vskip3mm

% Address

\address{$^1$Institute of Astronomy and National Astronomical Observatory, Bulgarian Academy of Sciences, 72, Tsarigradsko Shose Blvd., 1784 Sofia, Bulgaria}

\Email{esemkov}{astro.bas.bg, speneva@astro.bas.bg}

\address{$^2$Department of Physics and Astronomy, Faculty of Natural Sciences, University of Shumen, 115, Universitetska Str., 9712 Shumen, Bulgaria}

\Email{sibryamov}{shu.bg}

% Received and Accepted dates

\dates{.}{.}

% Abstract

\summary{We present results from photometric monitoring of V900 Mon, one of the newly discovered and still under-studied object from FU Orionis type.
FUor phenomenon is very rarely observed, but it is essential for stellar evolution.
Since we only know about twenty stars of this type, the study of each new object is very important for our knowledge.
Our data was obtained in the optical spectral region with BVRI Johnson-Cousins set of filters during the period from September 2011 to April 2021.
In order to follow the photometric history of the object, we measured its stellar magnitudes on the available plates from the Mikulski Archive for Space Telescopes.
The collected archival data suggests that the rise in brightness of V900 Mon began after January 1989 and the outburst goes so far.
In November 2009, when the outburst was registered, the star had already reached a level of brightness close to the current one.
Our observations indicate that during the period 2011-2017 the stellar magnitude increased gradually in each pass band.
The observed amplitude of the outburst is about 4 magnitudes (R).
During the last three years, the increase in brightness has stopped and there has even been a slight decline.
The comparison of the light curves of the known FUor objects shows that they are very diverse and are rarely repeated.
However, the photometric data we have so far shows that the V900 Mon's light curve is somewhat similar to this of V1515 Cyg and V733 Cep.}

% Keywords (see keywords.pdf file)


\begin{multicols}{2}

% Sections

\section{1. INTRODUCTION}

One of the main characteristics of the young stellar objects is their photometric variability.
In fact, most of the pre-main sequence (PMS) stars show variations in brightness that are associated with the evolutionary processes. 
PMS stars that undergo episodic outbursts with large amplitudes can be divided into two types FUors and EXors (Herbig 1989).
The common factor between the two types of eruptive phenomena is that they are observed on T Tauri stars.

The outburst of FU Orionis occurs in 1936 and for several decades this star remained the only object of this type.
First Ambartsumian (1971) draws attention to this object by linking it to the evolution of the PMS stars and proposes the abbreviation FUor.
After the discovery of two new FUor objects - V1057 Cyg and V1515 Cyg Herbig (1977) defined them as a new class of young eruptive stars.
Another dozen FUor objects were assigned to this class of young variables over the next four decades (Reipurth \& Aspin 2010, Audard et al. 2014, Connelley \& Reipurth 2018).
These objects have been classified in terms of their wide range of available photometric and spectral properties, but their outbursts are thought to have the same cause: an enhanced accretion rate from the circumstellar disk onto the central star (Hartmann \& Kenyon 1996, Herbig 2007).
During the outbursts, FUor objects undergo significant increase in their accretion rate from $\sim$10$^{-7}$$M_{\odot}$$/$yr up to $\sim$10$^{-4}$$M_{\odot}$$/$yr.

Several reasons are suggested to explain the enhanced accretion rate.
The most popular is that the increase is caused by thermal or gravitational instability in the circumstellar disk (Hartmann \& Kenyon 1996, Zhu et al. 2009).
Another possible triggering mechanisms could be the interactions of the circumstellar disk with a planet or nearby stellar companion on an eccentric orbit (Lodato \& Clarke 2004, Reipurth \& Aspin 2004, Pfalzner 2008) and in fall of clumps of material formed by disk fragmentation onto the central star (Vorobyov \& Basu 2005, Vorobyov et al. 2021).

Our experience in studying of FUor objects shows that the outburst can last for several decades or even a century.
As a rule, it is considered that the time to increase the brightness is much shorter than the time to decrease it.
All known FUors share the same defining characteristics: location in star-forming regions, association with reflection nebulae, a 4-6 mag. outburst amplitude, an F-G supergiant spectrum during the outbursts, a strong Li~I~6707~\AA\ line in absorption, and CO bands in near-infrared spectra (Herbig 1977, Reipurth \& Aspin 2010). 
An important feature of FUors is the massive supersonic wind observed as a P Cyg profile most commonly for both H$\alpha$ and Na I D lines.

The first three, also called classical FUor objects (FU Ori, V1515 Cyg and V1057 Cyg) are well-studied and their light curves are published in the literature (Clarke et al. 2005, Kopatskaya et al. 2013). 
Due to the large-scale optical and infrared monitoring programs carried out in several observatories and the contributions of amateur astronomers, some new objects have been observed to undergo outbursts with large amplitude: V733 Cep (Reipurht et al. 2007, Peneva et al. 2010), V2493 Cyg (Semkov et al. 2010, Semkov et al. 2012, Miller et al. 2011), V2492 Cyg (Aspin 2011, Hillenbrand et al. 2013, K{\'o}sp{\'a}l et al. 2013), V2494 Cyg (Aspin et al. 2009), V2495 Cyg (Movsessian et al. 2006), V582 Aur (Semkov et al. 2013, {\'A}brah{\'a}m et al. 2018), V2775 Ori (Fischer et al. 2012), V900 Mon (Reipurth et al. 2012), V960 Mon (Hillenbrand et al. 2014, K{\'o}sp{\'a}l et al. 2015,  Gaia 18bpy (Hillenbrand et al. 2018), Gaia 18dvy (Szegedi-Elek et al. 2020).
Due to the very small number of known FUor objects, each newly discovered attracts significant attention.

The detection of a new eruptive star in Monoceros (LDN 1656 cloud) has been reported by the amateur astronomer Jim Thommes (Thommes el. 2011).
According to the General Catalog of Variable Stars, the new variable is defined by the designation -- V900 Mon (Fig. 1).
Reipurth et al. (2012) performed detailed multi-wavelength study of V900 Mon and find significant similarities with the objects from the group of FUors. 
According to the authors the outburst of V900 Mon occurred between 1953 and 2009, and its luminosity is 106 $L_{\odot}$ at a distance of 1100 pc. 
Reipurth et al. (2012) also suggest that V900 Mon is a Class I source and the spectrum of the object is very similar to those of the classical prototype FU Orionis.
The spectrum of V900 Mon in the infrared region is much later than in the optical. 
Also the characteristic deep infrared CO bandhead absorption and the lithium Li~I~6707~\AA\ line, which is typical of young stars, are observed in case of V900 Mon. 

Gramajo et al. (2014) model the SED curve of V900 Mon in the 1–200 $\mu$m spectral range and estimate the basic parameters of the disk and the envelope of the star. 
The authors confirm that V900 Mon appears as a Class I source with disk mass 0.1 $M_{\odot}$ and disk mass accretion rate 2.0 $\times$ 10$^{-6}$ $M_{\odot}$ yr$^{-1}$.
They found considerable similarity to those of another eruptive star V1647 Ori after its outburst.
In order to study the mass and physical parameters of the envelope around V900 Mon K{\'o}sp{\'a}l et al. (2017) measured the $^{12}$CO and $^{13}$CO lines using the FLASH$^{+}$ receiver at the APEX telescope.
The results indicate that the mass of the envelope is low compared to other FUor objects and V900 Mon is at the end of the Class I evolutionary stage.

Varricatt et al. (2015) observed V900 Mon using the 3.8-m United Kingdom Infrared Telescope with the $L'$ and $M'$ MKO filters. 
The observations were made in April and September 2015 and show a slight increase in the brightness in the infrared region.
In our first paper concerning V900 Mon, we draw attention to the observed increase in brightness in the optical range during the period 2011-2016 (Semkov et al. 2017b). 
No other optical photometric data has been published since the outburst of V900 Mon was detected in 2009 and the object joined the FUors group.

Recently, data from observations of V900 Mon using ALMA array were published in two papers (Takami et al. 2019, Hales et al. 2020).
The presence of an extended molecular bipolar outflow and a rotating envelope is detected.
The disk around V900 Mon is resolved and its radius is set to $\sim$ 50 au.
(K{\'o}sp{\'a}l et al. 2020) note the presence of the 10 $\mu$m silicate feature in emission that is rather characteristic of Class II sources, but define it as a geometric effect from the high viewing angle.

\section{2. OBSERVATIONS AND DATA REDUCTION}

\subsection{2.1 CCD photometry}

The photometric observations of V900 Mon were performed with the 2 m RCC and the 50/70 cm Schmidt telescopes of the National Astronomical Observatory Rozhen (Bulgaria) and with the 1.3 m RC telescope of the Skinakas Observatory\footnote{Skinakas Observatory is a collaborative project of the University of Crete, the Foundation for Research and Technology - Hellas, and the Max-Planck-Institut f\"{u}r Extraterrestrische Physik.} of the Institute of Astronomy, University of Crete (Greece).  
Observations were performed with four types of CCD camera $-$ Vers Array 1300B (1340 $\times$ 1300 pixels, 20 $\times$ 20 $\mu m/$pixel size, scale 0.26 arcmin/pixel) and ANDOR iKon-L BEX2-DD (2048 $\times$ 2048 pixels, 13.5 $\times$ 13.5 $\mu m/$pixel size, scale 0.17 arcmin/pixel) at the 2-m RCC telescope, ANDOR DZ436-BV (2048 $\times$ 2048 pixels, 13.5 $\times$ 13.5 $\mu m/$pixel size, scale 0.33 arcmin/pixel) at the 1.3 m RC telescope and FLI PL16803 (4096 $\times$ 4096 pixels, 9 $\times$ 9 $\mu m/$pixel size, scale 1.08 arcmin/pixel) at the 50/70 cm Schmidt telescope.

Due to the negative Declination, V900 Mon can be observed from southern Europe only in the period September-April.
All frames were exposed through a set of standard Johnson-Cousins filters.
All the data was analyzed using the same aperture, which was chosen as 2 arcsec in radius, while the background annulus was taken from 9 arcsec to 14 arcsec in order to minimize the light from the surrounding nebula.

In order to facilitate the process of photometric measurement of the star, we calibrated twelve standard stars in close proximity to it.
Standard stars have a wide range of stellar magnitudes (from 13.689 to 17.286 mag in $I$-band).
In this way, we made it possible to compare current CCD observations with those from archival photographic plates.
Calibration was made during five clear nights in 2011 and 2012 with the 1.3 m RC telescope of the Skinakas Observatory and one night in 2011 with the 2-m RCC telescope of the Rozhen Observatory.
Standard stars from Landolt (1992) were used as a reference.  

\vskip.5cm
\centerline{\includegraphics[width=0.9\columnwidth,
keepaspectratio]{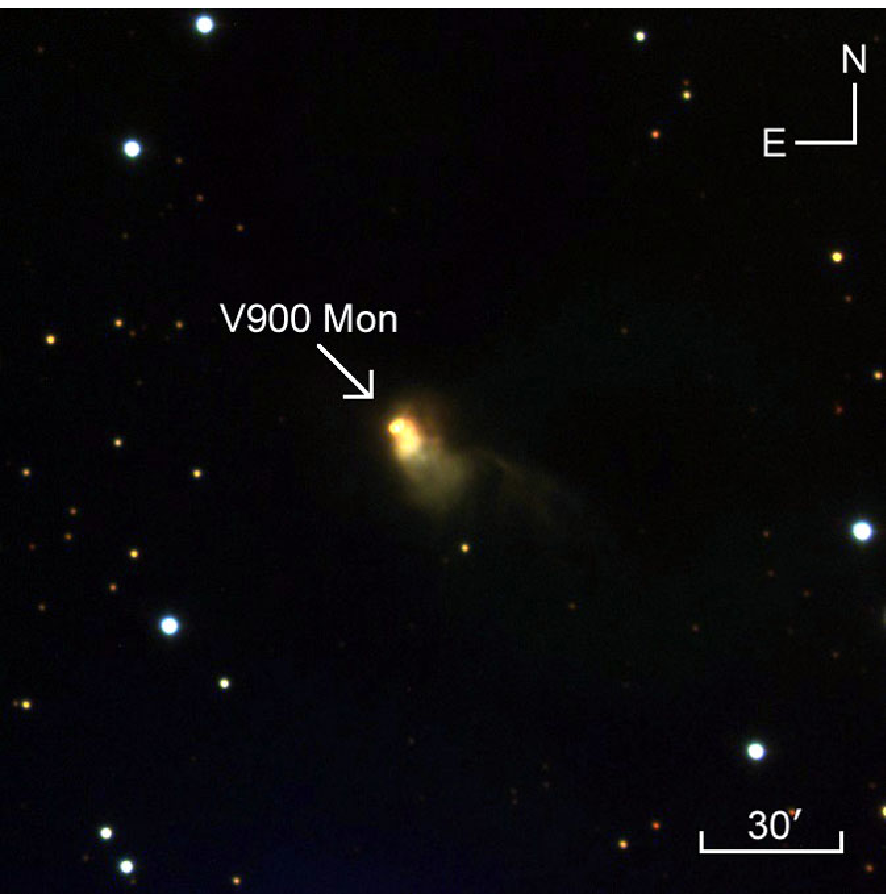}}

\figurecaption{1.}{Color image of V900 Mon obtained with the 2 m RCC telescope in NAO Rozhen. The cometary nebula arising from the star is clearly visible.}

Table 1 contains the $IRVB$ photometric data for the comparison sequence. 
The corresponding mean errors in the mean are also listed in the table. 
The measured stellar magnitudes of the standard stars in $B$-band are still uncertain and subject to improvement.
The stars are labeled from A to L in order of their $I$-band magnitude.  
In regions of star formation, a great percentage of stars can be photometric variables.  
Therefore, there is a possibility that some of our standard stars are low amplitude variables, and we advise observers to use our photometric sequence with care.  

The finding chart of the comparison sequence is presented in Fig. 2.  
The field is 8 $\times$ 8 arcmin, north is at the top and east is to the left.  
The chart is retrieved from the STScI Digitized Sky Survey Second Generation Red.

\end{multicols}

\noindent
\parbox{\textwidth}{
{\bf Table 1.} Photometric data for the $BVRI$ comparison sequence.
\vskip.25cm \centerline{
\begin{tabular}{lllllllll}
  
\hline\hline
\noalign{\smallskip}
Star &  $I$ & $\sigma_I$ & $R$ & $\sigma_R$ & $V$ & $\sigma_V$& $B$ & $\sigma_B$ \\
\noalign{\smallskip}
\hline
\noalign{\smallskip}
A &	13.689 &	0.019 &	14.099 &	0.019 &	14.551 &	0.014 &	15.3 & 0.1\\		
B &	13.789 &	0.025 &	14.183 &	0.023 &	14.572 &	0.014 &	15.2 & 0.1\\		
C &	14.471 &	0.024 &	15.084 &	0.015 &	14.821 &	0.014 &	16.8 & 0.1\\		
D &	14.541 &	0.021 &	15.951 &	0.019 &	17.258 &	0.029 &	19.4 & 0.3\\	
E &	14.902 &	0.026 &	15.525 &	0.018 &	16.141 &	0.018 &	17.1 & 0.2\\	
F &	15.948 &	0.022 &	16.575 &	0.021 &	17.185 &	0.046 &	18.1 & 0.2\\	
G &	15.953 &	0.040 &	16.480 &	0.021 &	17.167 &	0.034 &	18.7 & 0.2\\	
H &	15.991 &	0.077 &	16.764 &	0.022 &	17.536 &	0.015 &	$-$  & $-$\\		
I &	16.076 &	0.077 &	16.996 &	0.014 &	17.889 &	0.042 &	$-$  & $-$\\		
J &	16.079 &	0.025 &	16.989 &	0.036 &	17.726 &	0.036 &	18.7 & 0.2\\	
K &	16.469 &	0.055 &	17.465 &	0.031 &	18.267 &	0.028 & $-$  & $-$\\			
L &	17.286 &	0.084 &	18.485 &	0.029 &	19.324 &	0.105 & $-$  & $-$\\			
\hline                                  
\end{tabular}
} } 
\vskip.5cm

\begin{multicols}{2}

\vskip.5cm
	 \centerline{\includegraphics[width=0.9\columnwidth,
keepaspectratio]{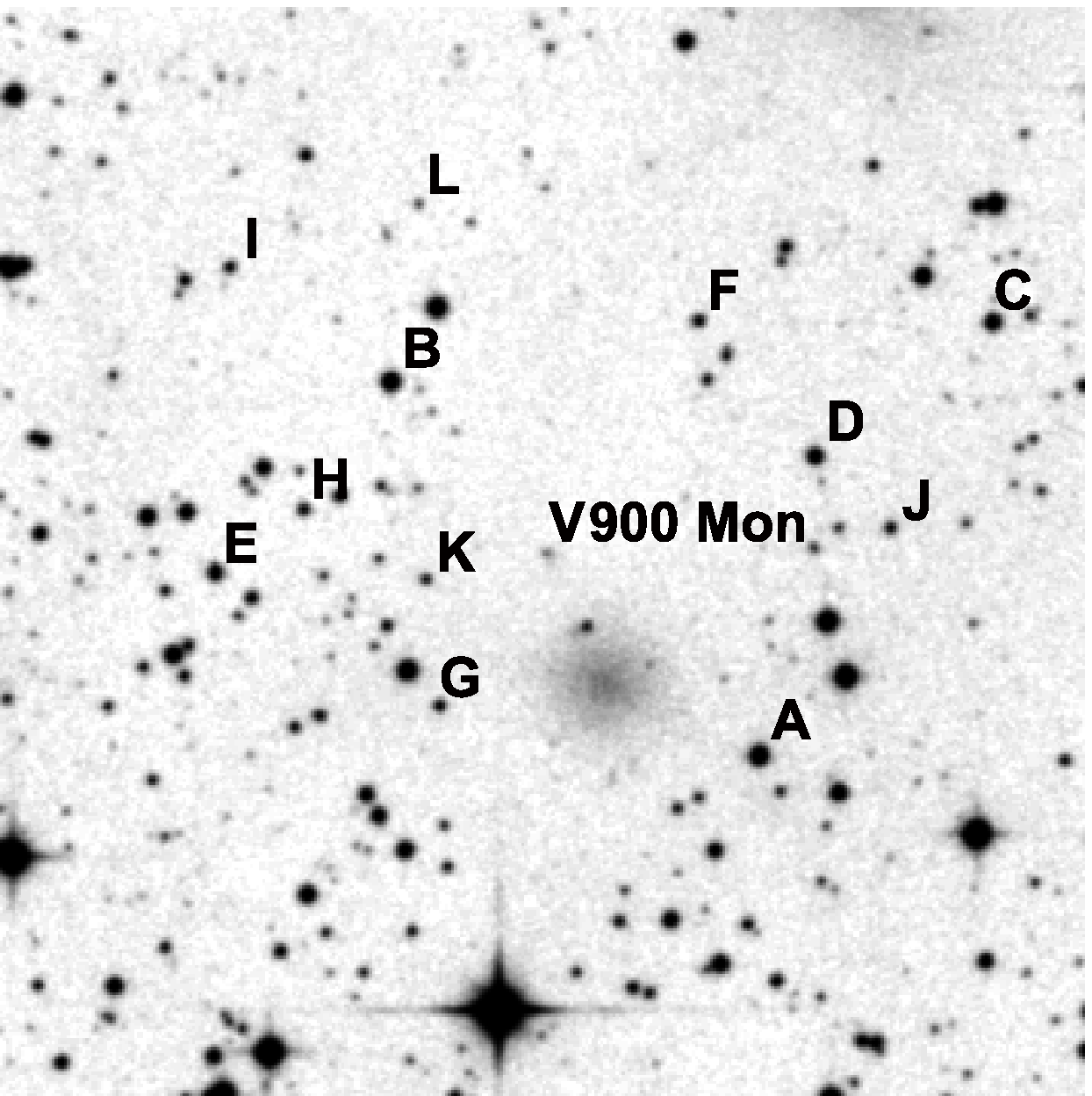}} 
	
   \figurecaption{2.}{Finding chart for the $\it BVRI$ comparison sequence around V900 Mon.}

\subsection{2.2 Data from photographic plates from the Mikulski Archive for Space Telescopes}

In order to investigate the photometric history of V900 Mon and to search for the beginning of the outburst, we searched the photographic plate archive of Mikulski Archive for Space Telescopes\footnote{$https://archive.stsci.edu/cgi-bin/dss$$\_$$plate$$\_$$finder$}.
As a result, we found scanned copies from a total of six photographic plates obtained with the Palomar Schmidt and the UK Schmidt telescopes.
The object was identified and measured on four of them. 
On the other two plates, the object is below the plate limit (XO661) and it is outside the field of the plate (S772).
Aperture photometry of the scanned plate copies were performed with DAOPHOT routines using the same aperture radius and the background annulus as for our CCD observations. 
In this way we achieve compliance of the measurements on the photographic plates with the CCD photometry.
The results of the measured magnitudes of V900 Mon from the scanned plate copies are given in Table 2.    

\end{multicols}

\noindent
\parbox{\textwidth}{
{\bf Table 2.} Photometric data from the Mikulski Archive for Space Telescopes.
\vskip.25cm \centerline{
\begin{tabular}{lllllll}
\hline\hline
\noalign{\smallskip}
Plate & Survey Name & Epoch & Emulsion+Filter & Exp.  & Plate Scale & Magnitude \\
ID    &             & yyyy-mm-dd hh:mm &      & min.       & arc. sec.  &            \\
\noalign{\smallskip}
\hline
\noalign{\smallskip}
ER700 & Equatorial Red   & 1989-01-08 14:19 &	IIIaF + OG590   & 63.0 & 1.01 &  R=18.60$\pm$0.04 \\
IS700 & SERC-I Survey    & 1985-02-10 10:56 & IVN + RG715     & 90.0 & 1.01 &  I=17.22$\pm$0.03 \\
S772  & SERC-J Survey    & 1983-01-17 12:25 & IIIaJ + GG395   & 60.0 & 1.70 &  out of plate \\
S700  & SERC-J Survey    & 1983-02-07 12:27 & IIIaJ + GG395   & 65.0 & 1.70 &  V=20.7$\pm$0.3 \\
XO661 & POSS-I O         & 1953-01-17 06:42 &	xx103aO         & 10.0 & 1.01 &  B$<$21 \\
XE661 & POSS-E Red       & 1953-01-17 07:52 &	xx103aE + plexi & 40.0 & 1.70 &  R=18.01$\pm$0.05\\
\hline 
\end{tabular}
} } 
\vskip.5cm
\begin{multicols}{2}

\section{3. RESULTS AND DISCUSSION}

The results of our photometric CCD observations of V900 Mon are summarized in Table 3.  
The columns provide the Julian Date of observation, $\it IRVB$ magnitudes of V900 Mon, the telescope and CCD camera used. 
The typical instrumental errors in the reported magnitudes are $0.01$-$0.02$ for $I$, $R$ and $V$-band, and $0.03-0.05$ for $B$-band.

The $BVRI$ light curves of V900 Cyg during the period of our observations are plotted in Fig. 3. 
The filled diamonds represent the CCD observations from the 50/70 cm Schmidt telescope, the filled squares CCD observations from the 2 m RCC telescope and the filled triangles the CCD observations from the 1.3 m RC telescope. 

\end{multicols}

\noindent
\parbox{\textwidth}{
{\bf Table 3.} Photometric $IRVB$ observations of V900 Mon during the period September 2011 - April 2021.
\vskip.25cm \centerline{
\begin{tabular}{llllllllllllll}
\hline\hline
\noalign{\smallskip}
J.D. (24...) & I & R & V & B & Tel & CCD & J.D. (24...) & I & R & V & B & Tel & CCD\\ 
\noalign{\smallskip}  
\hline
55815.612 &	13.30 & 14.94 &	16.11 &	      &	1.3RC &	ANDOR &	57369.520 &	13.02 &	14.55 &	15.54 &	16.74 &	2RCC  & VA\\
55816.627 &	13.32 & 14.95 &	      &	      &	1.3RC &	ANDOR &	57370.481 &	13.06 &	14.63 &	15.66 &	16.95 &	2RCC  & VA\\
55824.604 &	13.33 & 14.99 &	16.14 &	      &	1.3RC &	ANDOR &	57371.481 &	13.03 &	14.58 &	15.58 &	16.85 &	2RCC  & VA\\
55842.593 &	13.26 &	14.89 &	15.92 &       &	1.3RC &	ANDOR &	57372.489 &	13.05 &	14.58 &	15.42 &	      &	Sch   & FLI\\
55847.578 &	13.32 &	14.97 &	16.15 &     	&	1.3RC &	ANDOR &	57374.444 &	13.09 &	14.63 &	15.56 &	16.65 &	Sch   & FLI\\
55865.569 &	13.31 &	14.92 &	15.98 &	17.34 &	2RCC  &	VA &	57425.352 &	13.09 &	14.53 &	15.51 &	      &	Sch   & FLI\\
55866.594 &	13.32 &	14.93 &	16.02 &	17.33 &	2RCC  &	VA &	57426.341 &	13.12 &	14.60 &	15.77 &	      &	Sch   & FLI\\
55892.555 &	13.26 &	14.85 &	15.92 &	17.25 &	2RCC  &	VA &	57485.288 &	13.00 &	14.61 &	15.56 &	16.88 &	Sch   & FLI\\
55893.481 &	13.24 &	14.82 &	      &	      &	Sch   &	FLI &	57756.433 &	13.01 &	14.66 &	15.70 &	      &	Sch   & FLI\\
55895.547 &	13.33 &	14.80	&	      &	      &	Sch   &	FLI &	57781.348 &	12.96 &	14.53 &	15.52 &	16.54 &	Sch   & FLI\\
55896.502 &	13.33 &	14.79 &	      &	      &	Sch   &	FLI &	57781.429 &	12.99 &	14.38 &	15.71 &	16.91 &	2RCC  & VA\\
55925.505 &	13.16 &	14.65 &	      &	      &	Sch   &	FLI &	57785.361 &	12.97 &	14.39 &	15.63 &	16.99 &	2RCC  & VA\\
55928.422 &	13.20 &	14.74 &	15.90 &	      &	Sch   &	FLI &	57786.380 &	12.97 &	14.37 &	15.67 &	16.95 &	2RCC  & VA\\
56003.303 &	13.29 &	14.85 &	      &	      &	Sch   &	FLI &	57800.329 &	12.91 &	14.44 &	15.43 &	16.49 &	Sch   & FLI\\
56013.301 &	13.22 &	14.80 &	15.85 &	17.01 &	2RCC  &	VA &	57801.356 &	12.90 &	14.58 &	15.62 &	16.60 &	Sch   & FLI\\
56173.625 &	13.24 &	14.91 &	      &	      &	1.3RC &	ANDOR &	57813.305 &	12.90 &	14.52 &	15.53 &	      &	Sch   & FLI\\
56180.622 &	13.28 &	14.92 &	      &	      &	1.3RC &	ANDOR &	57817.312 &	12.80 &	14.15 &	15.20 &       &	Sch   & FLI\\
56183.627 &	13.29 &	14.92 &	16.08 &       &	1.3RC & ANDOR &	57845.236 &	13.01 &	14.62 &	15.60 &	      &	Sch   & FLI\\
56193.613 &	13.31 &	14.93 &	16.13 &	      &	1.3RC & ANDOR &	57846.298 &	13.00 &	14.48 &	15.61 &	      &	Sch   & FLI\\
56209.626 &	13.28 &	14.86 &	15.94 &	      &	Sch   & FLI &	58013.588 &	13.10 &	14.62 &	15.65 &	      &	Sch   & FLI\\
56249.563 &	13.36 &	14.97 &	16.04 &	      &	Sch   & FLI &	58043.588 &	13.02 &	14.60 &	15.60 &       &	Sch   & FLI\\
56275.512 &	13.14 &	14.66 &	15.65 &	      &	2RCC  & VA &	58080.536 &	12.98 &	14.57 &	15.52 &	16.66 &	Sch   & FLI\\
56276.505 &	13.26 &	14.82 &	15.89 &	17.24 &	2RCC  & VA &	58081.540 &	13.01 &	14.59 &	15.63 &	16.66 &	Sch   & FLI\\
56328.352 &	13.20 &	14.66 &	15.71 &	      &	Sch   & FLI &	58113.469 &	13.02 &	14.60 &	15.63 &	16.82 &	Sch   & FLI\\
56329.358 &	13.26 &	14.78 &	15.75 &	      &	Sch   & FLI &	58114.452 &	12.91 &	14.40 &	15.30 &	16.41 &	Sch   & FLI\\
56371.338 &	13.21 &	14.74 &	15.71 &	16.96 &	2RCC  & VA &	58217.268 &	12.95 &	14.52 &	15.50 &	      &	Sch   & FLI\\
56553.604 &	13.21 &	14.81 &	15.92 &	      &	1.3RC & ANDOR &	58218.260 &	12.98 &	14.56 &	15.72 &	      &	Sch   & FLI\\
56655.480 &	13.20 &	14.65 &	15.63 &	      &	Sch   & FLI &	58220.262 &	12.98 &	14.66 &	15.64 &	      &	Sch   & FLI\\
56656.445 &	13.23 &	14.86 &	15.81	&	      &	Sch   & FLI &	58220.281 &	13.13 &	14.46 &	15.72 &	17.08 &	2RCC  & ANDOR\\
56681.395 &	13.18 &	14.75 &	15.66 &	      &	Sch   & FLI &	58543.302 & 13.04 & 14.55 & 15.56 & 16.70 & Sch   & FLI\\
56694.362 &	13.19 &	14.78 &	15.83 &	17.21 &	2RCC  & VA &	58547.316 & 13.15 & 14.63 & 15.70 & 16.80 & Sch   & FLI\\
56715.289 &	13.19 &	14.65 &	15.50 &       &	Sch   & FLI & 58864.428 & 13.19 & 14.61 & 15.62 &       & Sch   & FLI\\
56738.253 &	13.13 &	14.62 &	15.53 &	      &	Sch   & FLI & 58865.437 & 13.01 & 14.36 & 15.59 &       & Sch   & FLI\\
57005.439 &	13.14 &	14.66 &	15.77 &	      &	Sch   & FLI & 58867.444 & 13.26 & 14.65 & 15.92 & 17.24 & 2RCC  & ANDOR\\
57006.523 &	13.16 &	14.69 &	15.75 &	      &	Sch   & FLI & 58869.420 & 13.11 & 14.46 & 15.61 & 16.71 & 2RCC  & ANDOR\\
57016.486 &	13.16 &	14.69 &	15.64 &	16.91 &	2RCC  & VA &  58870.429 & 13.24 & 14.74 & 15.81 &       & Sch   & FLI\\
57072.314 &	13.05 &	14.42 &	15.23 &	      &	Sch   & FLI & 58901.330 & 13.05 & 14.44 & 15.44 &       & Sch   & FLI\\
57074.328 &	13.18 &	14.69 &	15.67 &	      &	Sch   & FLI & 58930.281 & 13.15 & 14.60 & 15.75 &       & Sch   & FLI\\
57268.618 &	13.12 &	14.70 &	15.80 &	      &	1.3RC & ANDOR & 59176.505 & 13.19 & 14.54 & 15.61 & 16.69 & Sch & FLI\\
57269.622 &	13.10 &	14.69 &	15.74 &	      &	1.3RC & ANDOR & 59177.516 & 13.09 & 14.43 & 15.42 & 16.58 & Sch & FLI\\	
57330.548 &	13.07 &	14.60 &	15.51 &	      &	Sch   & FLI & 59220.388 & 12.96 & 14.35 & 15.50 & 16.87 & 2RCC & ANDOR\\
57331.561 &	13.08 &	14.56 &	15.49 &	      &	Sch   & FLI & 59250.392 & 13.10 & 14.43 & 15.53  &     & 2RCC & ANDOR\\
57333.594 &	13.11 &	14.65 &	15.61 &	      &	Sch   & FLI & 59251.337 & 13.12 & 14.45 & 15.64 & 16.86 & 2RCC & ANDOR\\
57714.551 &	13.02 &	14.47 &	15.70 &	16.99 &	2RCC  & VA &  59314.253 & 12.94 & 14.30 & 15.40 & 16.45 & 2RCC & ANDOR\\
57715.544 &	13.06 &	14.47 &	15.72 &	17.05 &	2RCC  & VA &  59316.258 & 12.93 & 14.32 & 15.33 & 16.54 & Sch & FLI\\
57716.536 &	13.06 &	14.48 &	15.70 &	17.02 &	2RCC  & VA\\
\hline
\end{tabular}
} } 
\vskip.5cm

\begin{multicols}{2}

\end{multicols}
 \begingroup\centering
	 \includegraphics[width=16cm, angle=0]{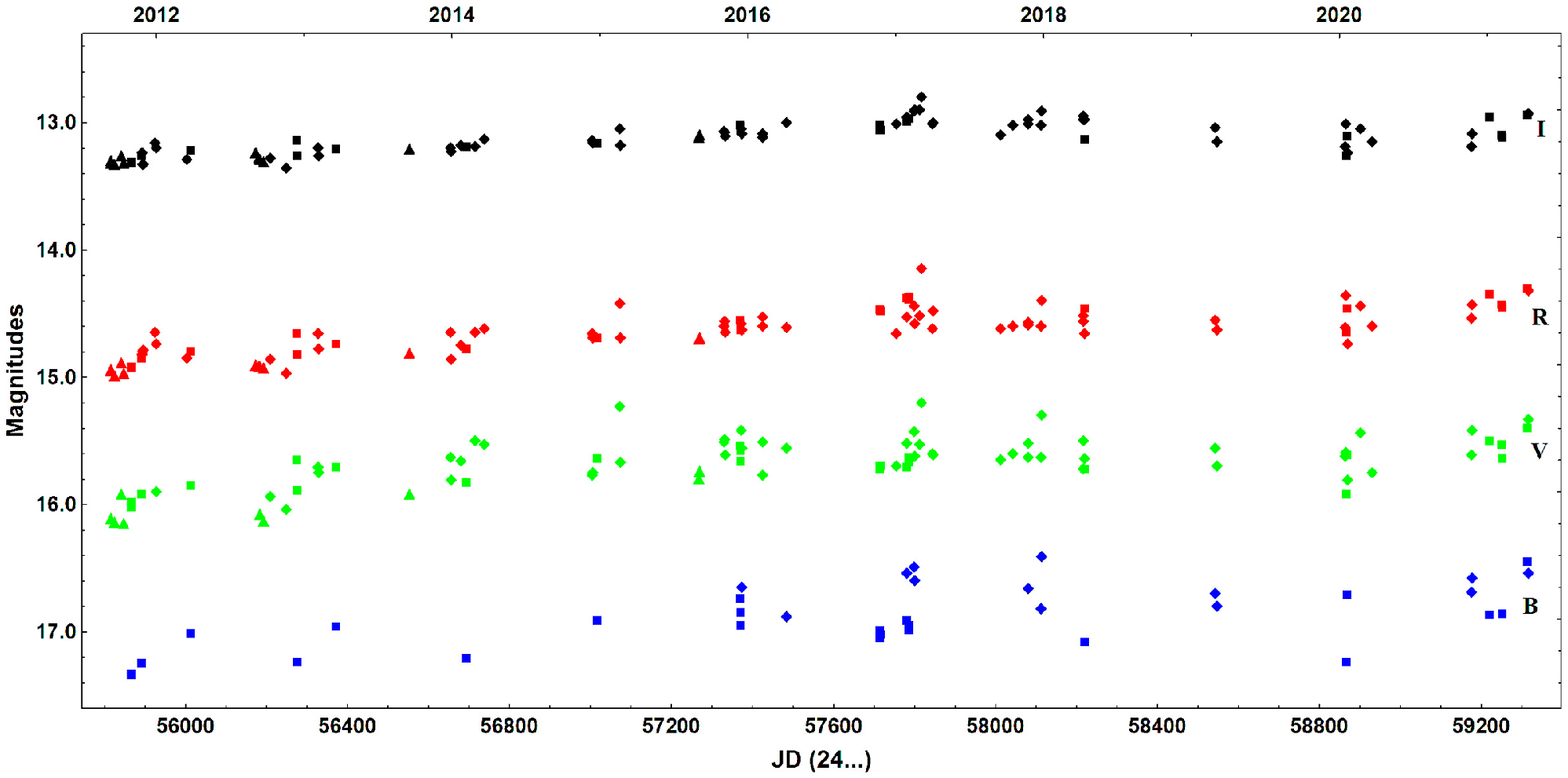}
   \figurecaption{3.}{BVRI light curves of V900 Mon for the period of our observations 2011 September $ - $ 2021 April.}
   \endgroup
\begin{multicols}{2}

The historical $VRI$ light curves of V900 Mon from all available observations are plotted in Fig. 4. 
Our CCD observations are indicated by the same symbols as on Fig. 3.
The filled circles on Fig. 4 represent data from Mikulski Archive for Space Telescopes, listed in this paper.
The empty circles represent data from DSS reported in Table 1 of the paper of Reipurth et al. (2012).

Archival data shows that before the outburst V900 Mon was variable, which is especially noticeable in the $R$-light.
The amplitude of variability is about 1 magnitude, which is a typical variability of T Tauri stars.
There are several FUor objects for which the data for photometric variability before the outburst is available (FU Ori, V1057 Cyg, V2493 Cyg, V582 Aur, Gaia 18dvy).
They all show non-periodic changes in brightness, that proves the affiliation of the prototypes of FUors to the class of T Tauri stars.

The exact time of the beginning of the outburst of V900 Mon cannot be determined now.
But it can certainly be concluded that the outburst occurred after January 1989.
Due to the lack of data for the period 1989-2011, we cannot determine the rate of increase in brightness and the nature of the light curve before reaching the maximum brightness.
The first photometric data for V900 Mon from 2009 and 2010 reported by Thommes el. (2011) and Reipurth et al. (2012) were obtained in the $g'r'i'z'$ system and are not compatible with our results.
Also, this data was obtained with a smaller aperture (1 arcsec), and as noted by Reipurth et al. (2012), the aperture radius is of great importance for the measured value of the stellar magnitudes.

Since the beginning of our observations in 2011, we have been recording a gradual rise in brightness.
But during the last three observation seasons (from the end of 2018 to the beginning of 2021), the increase in brightness has stopped and there has even been a slight decline.
The increase in brightness is noticeable in all filters, as for the R-band it is about 0.5 mag. for the eight-year period from 2011 to 2017.
Therefore, V900 Mon has reached its maximum brightness during observation season 2017$-$2018 and we should expect a decrease in brightness in the coming years.
But occasionally it is possible to observe two peaks at maximum brightness, as is the case of V2493 Cyg outburst (Semkov et al. 2012). 
Whether we have really registered the maximum of the FUor type outburst should be confirmed by subsequent observations.

Long-term photometric data from observations of known FUor objects show a wide variety of the light curves.
Even the first three classical FUor objects show different rates of increase and decrease in brightness (Clarke et al. 2005).
As the number of known FUor objects increases, the diversity of the light curves increases even more.
Some objects show a very rapid rise in brightness within a few months or a year (FU Ori, V1057 Cyg, V2493 Cyg) (Clarke et al. 2005, Semkov et al. 2012). 
In others, the rise in brightness can take several years and even achieves to 20-30 years period (V1515 Cyg, V1735 Cyg, V733 Cep) (Clarke et al. 2005, Peneva et al. 2009, Peneva et al. 2010). 

Such diversity is observed for the periods of decline in brightness.
Usually the decline in brightness lasts for decades or even a century. 
But there are objects where a comparatively rapid decline in brightness has been observed. 
For example, V960 Mon in which the brightness decreases by 2 mag (V) over a period of five years (Takagi et al. 2020), 
and V582 Aur for which two deep decreases in brightens by about 3 mag (R) have been observed separated by a five-year period (Semkov et al. 2013, {\'A}brah{\'a}m et al. 2018).
There are also objects that for a period of several decades do not practically change their brightness as in the case of V1735 Cyg (Peneva et al. 2009).
Also, V2493 Cyg remains for several years with an almost constant magnitude at the maximum level of brightness (Semkov et al. 2017a).

This variety of photometric behavior strongly supports the idea that FUor objects are not a homogeneous group and that the causes of this phenomenon can be several different mechanisms (Vorobyov et al. 2021). 
Trying to compare the available photometric data for V900 Mon with the other FUors, we register the following similarities. 
The slow rise in brightness during the recent years is similar to that of objects V1515 Cyg and V733 Cep. 
We do not yet have data on how the decline in brightness will continue, but the beginning of a gradual decline is also similar to the above two objects.

   \end{multicols}
   \begingroup\centering
   \includegraphics[width=16cm, angle=0]{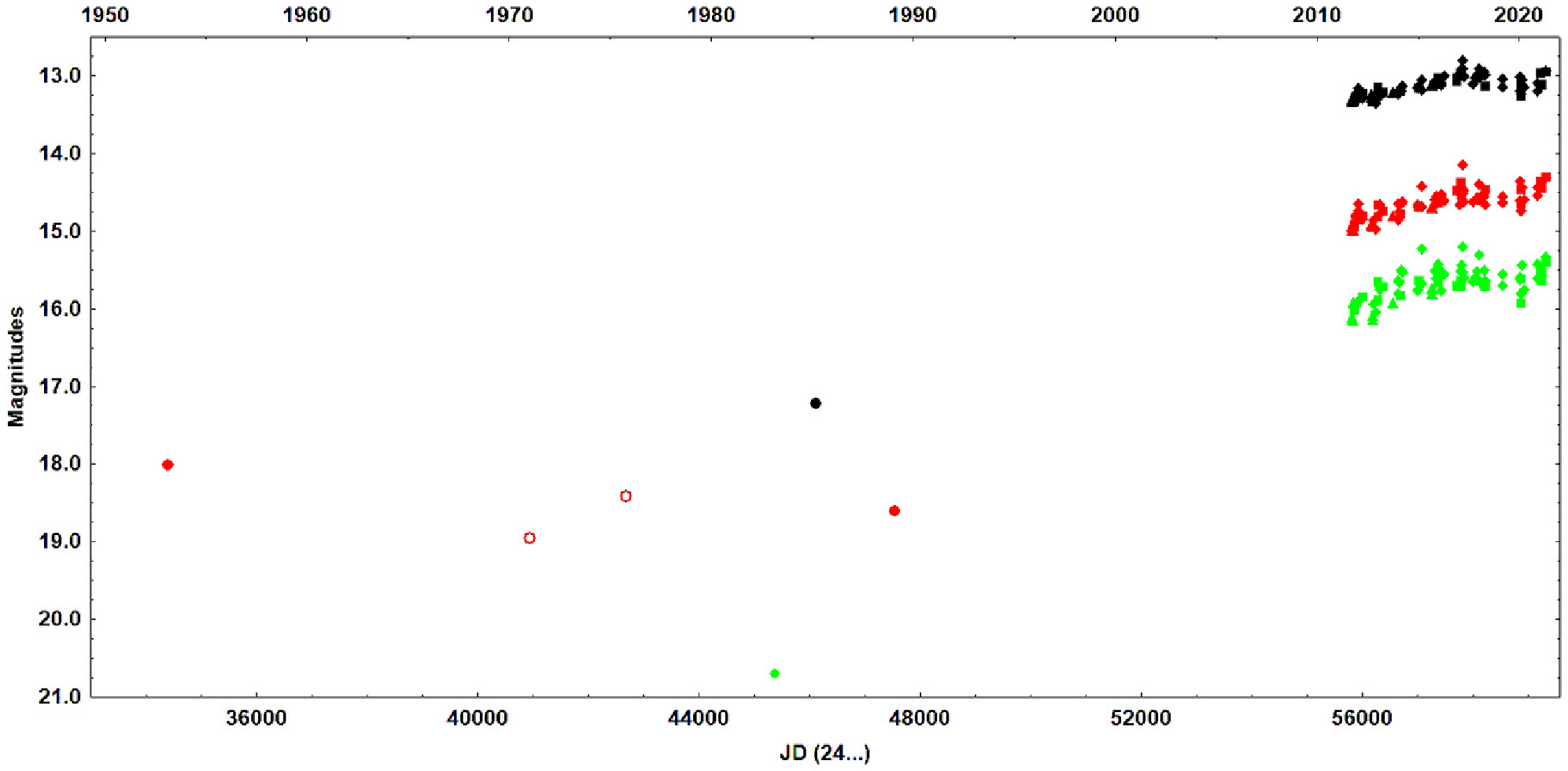}
   \figurecaption{4.}{Historical VRI light curves of V900 Mon for the period 1953 January$ - $2021 April.
In the figure, the data from the $I$ pass band are colored in black, from the $R$ pass band in red and from the $V$ pass band in green.}
   \endgroup
		\begin{multicols}{2}

\section{4. CONCLUSIONS}

During the last twenty years, we have undertaken intensive photometric monitoring of several unexplored FUor objects.
This data can also be used to detect new FUor or EXor events and to determine the type of the outburst. 
FUors are the most noticeable young variable objects due to the large amplitudes and eruptions lasting for years.
Due to the small number of such objects, we do not know enough about their photometric history. 

The data collected so far for V900 Mon shows that it can be classified with confidence as a $bona fide$ FUor.
An outburst with an amplitude of 4 magnitudes was registered, which began after 1989. 
Maximum brightness was observed in the period 2017-2018, after which a decrease in brightness began. 

Our results show the need for systematical photometric monitoring of the objects of special interest. 
Collecting more data from observations as well as from astronomical archives can help clarify theories about the early stages of stellar evolution. 
We support the idea that the type of light curve of a FUor object can be evidence of the specific cause of its outburst.

% Acknowledgements

\acknowledgements{This work was partly supported by the Bulgarian Scientific Research Fund of the Ministry of Education and Science under the grant DN 18-13/2017 and DN 18-10/2017. 
The authors thank the Director of Skinakas Observatory Prof. I. Papamastorakis and Prof. I. Papadakis for the award of telescope time. 
This research has made use of the NASA's Astrophysics Data System Abstract Service, the SIMBAD database and the VizieR catalogue access tool, operated at CDS, Strasbourg, France. 
Some of the data presented in this paper was obtained from the Mikulski Archive for Space Telescopes (MAST). 
STScI is operated by the Association of Universities for Research in Astronomy, Inc., under NASA contract NAS5-26555.}

% References

\references

{\'A}brah{\'a}m, P., K{\'o}sp{\'a}l, {\'A}., Kun, M., et al. 2018, \journal{Astrophys. J.}, \vol{853}, 28

Ambartsumian, V. A. 1971, \journal{Astrophysics}, \vol{7}, 331

Aspin, C. 2011, \journal{Astron. J.}, \vol{142}, 135

Aspin, C., Beck, T. L., Pyo, T.-S., et al. 2009, \journal{Astron. J.}, \vol{137}, 431

Audard, M., {\'A}brah{\'a}m, P., Dunham, M. M., et al. 2014, in Protostars and Planets VI, ed. H. Beuther et al. (Tucson, AZ: Univ. Arizona Press), p.~387

Clarke, C., Lodato, G., Melnikov, S. Y., \& Ibrahimov, M. A. 2005, \journal{Mon. Not. R. Astron. Soc.}, \vol{361}, 942

Connelley, M. S., \& Reipurth, B. 2018, \journal{Astrophys. J.}, \vol{861}, 145

Fischer, W. J., Megeath, S. T., Tobin, J. J., et al. 2012, \journal{Astrophys. J.}, \vol{756}, 99

Gramajo, L. V., Rod{\'o}n, J. A., \& G{\'o}mez, M. 2014, \journal{Astron. J.}, \vol{147}, 140

Hales, A. S., P{\'e}rez, S., Gonzalez-Ruilova, C. et al. 2020, \journal{Astrophys. J.}, \vol{900}, 7

Hartmann, L., \& Kenyon, S. J. 1996, \journal{Annu. Rev. Astron. Astrophys.}, \vol{34}, 207

Herbig, G. H. 1977, \journal{Astrophys. J.}, \vol{217}, 693

Herbig, G. H.: 1989, in ESO Workshop on Low-Mass Star Formation and Pre-Main-Sequence Objects, ed. B. Reipurth (Garching: ESO), 233

Herbig, G. H.: 2007, \journal{Astrophys. J.}, \vol{133}, 2679 

Hillenbrand, L. A., Miller, A. A., Covey, K. R. et al. 2013, \journal{Astron. J.}, \vol{145}, 59
  
Hillenbrand, L. 2014, \journal{The Astronomer’s Telegram}, 6797

Hillenbrand, L. A., Contreras Pe{\~n}a, C., Morrell, S. et al. 2018, \journal{Astrophys. J.}, \vol{869}, 146

Kopatskaya, E. N., Kolotilov, E. A., \& Arkharov, A. A. 2013, \journal{Mon. Not. R. Astron. Soc.}, \vol{434}, 38

K{\'o}sp{\'a}l, {\'A}., {\'A}brah{\'a}m, P., Acosta-Pulido, J. A. et al. 2013, \journal{Astron. Astrophys}, \vol{551}, A62

K{\'o}sp{\'a}l, {\'A}., {\'A}brah{\'a}m, P., Mo{\'o}r, A., et al. 2015, \journal{Astrophys. J. Lett.}, \vol{801}, L5

K{\'o}sp{\'a}l, {\'A}., {\'A}brah{\'a}m, P., Csengeri, T. et al. 2017, \journal{Astrophys. J.}, \vol{843}, 45

K{\'o}sp{\'a}l, {\'A}., {\'A}brah{\'a}m, P., Carmona, A. et al. 2020, \journal{Astrophys. J. Lett.}, \vol{898}, L48

Landolt, A. U. 1992, \journal{Astron. J.}, \vol{104}, 340

Lodato, G., \& Clarke, C. J.: 2004, \journal{Mon. Not. R. Astron. Soc.}, \vol{353}, 841

Miller, A, A., Hillenbrand, L. A., Covey, K. R. et al. 2011, \journal{Astrophys. J.}, \vol{730}, 80

Movsessian, T. A., Khanzadyan, T., Aspin, C., et al. 2006, \journal{Astron. Astrophys}, \vol{455}, 1001

Peneva, S. P., Semkov, E. H., \& Stavrev, K. Y. 2009, \journal{Astrophys. Space Sci.}, \vol{323}, 329

Peneva, S. P., Semkov, E. H., Munari, U., \& Birkle, K. 2010, \journal{Astron. Astrophys}, \vol{515}, A24

Pfalzner, S.: 2008, \journal{Astron. Astrophys}, \vol{492}, 735

Reipurth, B., \& Aspin, C.: 2004, \journal{Astrophys. J. Lett.}, \vol{608}, L65

Reipurth, B., Aspin, C., Beck, T. et al. 2007, \journal{Astron. J.}, \vol{133}, 1000

Reipurth, B., \& Aspin, C., 2010, in Evolution of Cosmic Objects through their Physical
Activity, eds. H. A. Harutyunian, A. M. Mickaelian, Y. Terzian (Yerevan: Gitutyun), p.~19

Reipurth, B., Aspin, C., \& Herbig, G. H. 2012, \journal{Astrophys. J. Lett.}, \vol{748}, L5

Semkov, E. H., Peneva, S. P., Munari, U., Milani, A., \& Valisa, P. 2010, \journal{Astron. Astrophys}, \vol{523}, L3 
  
Semkov, E. H., Peneva, S. P., Munari, U. et al. 2012, \journal{Astron. Astrophys}, \vol{542}, A43
  
Semkov, E. H., Peneva, S. P., Munari, U., Dennefeld, M., Mito, H., Dimitrov, D. P., Ibryamov, S., \& Stoyanov, K. A. 2013, \journal{Astron. Astrophys}, \vol{556}, A60

Semkov, E. H., Peneva, S. P., \& Ibryamov, S. 2017a, \journal{Bulg. Astron. J.}, \vol{26}, 57

Semkov, E., Peneva, S., \& Ibryamov, S. 2017b, in IAU Symp. 325, Astroinformatics, ed. M. Brescia et al. (Cambridge: Cambridge Univ. Press), 266

Szegedi-Elek, E., {\'A}brah{\'a}m, P., Wyrzykowski, Ł. et al. 2020, \journal{Astrophys. J.}, \vol{899}, 130 

Takagi, Y., Honda, S., Arai, A., et al. 2020, \journal{Astrophys. J.}, \vol{904}, 53 

Takami, M., Chen, T.-S., Liu, H. B., et al. 2019, \journal{Astrophys. J.}, \vol{884}, 146

Thommes, J., Reipurth, B., Aspin, C., \& Herbig, G. H. 2011, CBET, 2795, 1

Varricatt, W.P., Kerr, T.H., Carroll, T., \& Moore, E. 2015, ATel, 8174, 1 

Vorobyov, E. I., \& Basu, S. 2005, \journal{Astrophys. J. Lett.}, \vol{633}, L137

Vorobyov, E. I., Elbakyan, V. G., Liu, H. B., \& Takami, M. 2021, \journal{Astron. Astrophys}, \vol{647}, A44

Zhu, Z., Hartmann, L., Gammie, C., \& McKinney, J. C. 2009, \journal{Astrophys. J.}, \vol{701}, 620

\endreferences

%\end{multicols}

\end{multicols}
\end{document}